\documentclass[nohyper,12pt,letterpaper]{JHEP}
\usepackage{epsfig}

\newfont{\frak}{eufm10 scaled 1200}

\newfont{\Bbb}{msbm10 scaled 1200}     
\newcommand{\mathbb}[1]{\mbox{\Bbb #1}}
\DeclareSymbolFont{AMSa}{U}{msa}{m}{n}
\DeclareSymbolFont{AMSb}{U}{msb}{m}{n}
\let\Box\relax
\DeclareMathSymbol{\Box}{\mathord}{AMSa}{"03}

\def \eqn#1#2{\begin{equation}#2\label{#1}\end{equation}}

\title{On Isolated Vacua and Background Independence}

\author{T. Banks\thanks{On leave from U.C. Santa Cruz.}\\
  Department of Physics and Astronomy\\
  Rutgers University, Piscataway, NJ 08855-0849\\
  E-mail: \email{banks@physics.rutgers.edu}}

\abstract{I argue that isolated vacua of M-theory, cannot in any 
conventional way be said to live in the same
theory as other disconnected parts of the moduli space.  
The usual field theoretic mechanisms, which allow an observer in one 
disconnected component 
of a moduli space to verify the existence of other components, fail.
The failure is a consequence of robust properties of black holes.  When
barriers between components are much smaller than the Planck scale, the
usual field theoretic picture is approximately valid.}

\keywords{M-Theory, Background Independence}

\received{???????? ?st, 2000} \accepted{???????? ?th, 2000}

\preprint{\hepth{0011255}\\RUNHETC-2000-51\\SCIPP-00/41}
\begin{document}

\section{\bf Introduction - a vacuous diatribe}

There has been much speculation about the topology of the space of
asymptotically flat M-theory vacua, by which I mean quantum mechanical 
states of the system
with some number of asymptotically flat dimensions with propagating
gravitons and exact Poincare symmetry.   In cases where eight or more
supercharges are also included in the asymptotic symmetry group, the
situation is, as I review briefly below, fairly well understood.  There
are continuous moduli spaces of such vacua, which can reasonably
reliably be said to constitute states of ``the same theory'' in a sense
that I will make more precise. The main purpose of this article is to
investigate whether the same can be said of vacua with only four
supercharges, which perforce have only four asymptotically flat
dimensions. Our considerations will be more general, and actually apply
to any two disconnected components of moduli space, but for the most
part I will use the language of theories with four supercharges in four
dimensions.

In this case, continuous moduli spaces are not generic.   Indeed, given
the absence of exact continuous global symmetries in string theory, one
expects a superpotential of general form on the would be moduli space
and this implies that SUSY vacua form a discrete set.  Indeed, if we
impose the extra restriction of vanishing cosmological constant, the
equations are overdetermined and should only be satisfied at points of
enhanced discrete complex R-symmetry.   It is only if some of the
(infinitesimal variations) of the moduli at such an enhanced symmetry
point carry vanishing discrete R charge that we can expect a moduli
space of nonvanishing dimension \cite{bdcoping}.  Within such $N=1$
moduli spaces, the arguments alluded to in the first paragraph go
through and the different vacua are states of the same theory.  If these
moduli spaces contain boundaries where space decompactifies and an
enhanced SUSY algebra is restored, then these vacua are likely to be in
the same theory as the higher dimensional, 
extended SUSY vacua referred to above.

Now however, consider an isolated vacuum with four
supercharges\footnote{Our considerations apply equally well to any two
disconnected components of moduli space with vanishing cosmological
constant.} .  
In local quantum field theory we have no problem asserting that such a
vacuum state might be in the same theory as any other isolated SUSY
vacuum, or the continuous moduli spaces referred to above.  Indeed,
consider a field theory with two degenerate minima of the potential.
If we are willing to expend enough energy, we can easily construct an
excited state of the theory in one of these vacuum states, that contains
an arbitrarily large region which looks like the other.   This is a
bubble and its energy scales as the square of its radius.  For large
radius it is unstable, and shrinks, but causality guarantees that such a
large bubble cannot dissipate for times much less than $R$.   Thus, an
experimentalist with a large amount of energy at his disposal can create
and study such a vacuum bubble.

In this paper we will argue that the same is not true in a theory
containing gravity.  At its crudest, the argument consists of the simple
observation that if $\sigma$ is the surface tension of the bubble, then
the Schwarzchild radius of the large distance gravitational field of the
bubble is $\sigma (R/m_P)^2$.  Thus, the bubble will be smaller than its
Schwarzchild radius if $R > m_P^2 / \sigma$.  Thus, unless $\sigma$ is
very tiny in Planck units, an experimenter who attempts to create a
large bubble of an alternative universe will instead create a black
hole.  This observation is closely related to previous studies of
``creating a universe in the laboratory'' \cite{gf} and more
particularly to the seminal work of Cvetic and collaborators
\cite{cvetic} on domain walls between SUSY vacua.  We will give a more
precise discussion of this work in the next section.

For the moment however, let us return to exact moduli spaces of vacua.
In this context, the difficulty mentioned in the previous paragraph does
not arise.  There is no potential energy on the space of vacua, and we
can construct a configuration with a large region of another vacuum at
an arbitrarily small cost in energy if we are willing to let the wall
interpolating between the bubble and the outside world be very thick (as
we will see, this is not an option if we have a
potential)\footnote{Conversations with Barak Kol, L. Motl and
C. Roemelsberger have shown that this discussion is more intricate than
one would have thought.  In order to explore regions of moduli space
which are of order $N M_P$ apart (in four dimensional normalization for
canonical fields) we must construct bubbles which grow exponentially
with $N$ and then have a series of small bubbles inside them.  The step
in field space between a consecutive pair of bubbles is bounded by
something of order the Planck scale. The wall of each bubble is about
as thick as its radius.}.
Thus, points at finite separation in exact moduli spaces should surely
be thought of as being in the same theory.

Nonetheless, the situation is, even in this case, not quite what SUSY
field theory intuition would lead us to believe.   The problem is again
associated with black holes, though in a superficially different manner.
In non-gravitational quantum mechanical models there are two related
notions of what it means for two states to be in the same theory.  The
first comes from classical dynamics, and conditions most of our thinking
about these problems.  One envisages a classical system with dynamical
variables $q^i (t)$ and considers two static solutions to be in the same
theory because they correspond to different solutions of the same
equations of motion.  When the system is a field theory in infinite
volume, one observes that if the two solutions are also constant in
space, then the solution space breaks up into classes which can be
viewed as local perturbations of one or the other classical vacuum.

More abstractly, and more generally, one recalls that quantum field
theories are defined by (perhaps marginally) relevant perturbations of a
fixed point conformal field theory.  By studying correlation functions
at small Euclidean separations we find the physics dominated by the
fixed point and the relevant perturbations of it.  In this way, we can 
determine that two different vacua of the same theory are really in the
same theory.  The superselection rule that prevents us from putting
both vacua in the same Hilbert space has a purely infrared character,
and depends on the notion of locality.  The Hilbert space obtained by
applying local fields to one vacuum state is orthogonal to that obtained
by applying them to the other.  By studying short distance correlation
functions, we can identify the theory underlying each vacuum and verify
that they both belong to the same field theory (the same fixed point
perturbed by the same relevant operators).

It is worth pointing out that the classical notion of when two vacua
come from the same theory is a special case in which the fixed point
theory is Gaussian.  Indeed, Feynman's derivation of the path integral
formalism proceeds by concatenating propagation over infinitely short
time intervals.  The Lagrangian formalism emerges when one makes the
assumption that the short time behavior is dominated by a quadratic
theory. Thus, in the quantum theories we understand, the way to test
whether two ground states are states of the same theory is to examine
short time, or high energy behavior.

In Asymptotically Flat M (AFM) Theory\footnote{As always, I insist on
using the term M-theory to describe the unified theory from which weakly
coupled string and SUGRA perturbation expansions are derived as limits.
It is, at the very least, uneconomical to reserve this term for the
region of moduli space near the 11 dimensional SUGRA limit.} it was
pointed out in \cite{banksah} and \cite{bfbh} that the physics of
asymptotically high energies is dominated by black holes.  This is an
example of the UV-IR \cite{sw} connection in M-theory.  High energy
states are associated with large objects with low curvatures and large
degeneracies. Some aspects of the physics of these objects is
consequently accessible to long wavelength SUGRA analysis.  The black
hole density of states is unlike anything in quantum field theory.  It
implies that operators which cannot resolve the black hole degeneracy
cannot have well defined correlation functions.  More precisely ({\it
cf. } \cite{kap}) generic Heisenberg operators $O(t)$ 
in such a theory are operator valued 
distributions on a space of test functions which cannot be localized in
time (their Fourier transforms must fall off so rapidly that they cannot
vanish outside compact intervals in time).

For our present purposes, what is most important is that the black hole
density of states is dimension dependent.  Thus, {\it two AFM vacua with
different numbers of AF dimensions are not vacua of the same theory in
the sense with which we are familiar in Quantum Field Theory}.  The
relation between them is more subtle.  The maximum growth of the density
of states occurs for four noncompact spacetime dimensions.  If one has a
moduli space of such four dimensional compactifications, one can find
boundaries of moduli space in which more noncompact dimensions appear.

As we take this decompactification limit, states of extended objects
wrapped around cycles of the compact manifold will go off to infinite
energy.  It is no surprise to an M-theorist then (but would be to a
quantum field theorist) that if we take the high energy limit after
decompactification, we get a different behavior of the density of states
than we did before.  The relationship between moduli spaces of AFM
theory with different numbers of noncompact spacetime dimensions but the
same number of SUSYs can then be understood in this limiting fashion.
All compactifications to four dimensions with a fixed number of SUSY
generators, which lie on the same continuous moduli space, are plausibly
states of the same theory.  States with more noncompact dimensions can
be obtained as limits of these.  The limiting process is singular and
does not commute with the high energy limit.  A question left unanswered
by this analysis is the relation between moduli spaces of vacua with the
same number of AF dimensions, but different numbers of SUSYs.

The case of only four supercharges deserves a separate discussion.
Here, general effective field theory arguments \cite{bdcoping} suggest
that most SUSY vacua will have negative cosmological constant.
Points with enhanced complex R-symmetry can naturally have vanishing
cosmological constant but will generically be isolated.  Only if there
are chiral superfields with vanishing R-charge can one expect continuous
moduli spaces.   The moduli space thus splits into disconnected
components, most isolated points, but perhaps some of  nonzero
dimension. The aim of the rest of this paper is to study the extent to
which we can consider these disconnected components as part of the same
theory.

\section{\bf Walls of evidence}

The first clue that isolated vacua cannot generically communicate with
each other, comes from an examination of hypothetical domain walls
separating two such vacua.  In field theory, the solutions in which an
observer in one vacuum injects energy and creates a large, long-lived 
bubble of another, asymptote, as the energy goes to infinity, to static,
stable domain wall solutions.  Fortunately, in a beautiful series of
papers some years ago \cite{cvetic}, Cvetic and collaborators made a
comprehensive study of domain walls in supergravity, in a variety of
situations.  Here I will only give a partial summary of their work, and
I urge the reader to consult the original papers for more detail and for
the description of situations with dilaton fields and/or AdS vacua,
which are not relevant to the present paper.

It is well known that SUGRA can have supersymmetric vacuum states with
nonpositive values of the cosmological constant.  The value of the
cosmological constant is determined by the value of the superpotential
at the supersymmetric points.  There are static, stable BPS domain walls
between two such vacua with different values of the cosmological
constant.  The tension of the BPS walls is proportional to the
difference of superpotentials in the two vacua.

By examining solutions with the symmetries of domain wall spacetimes in
the thin wall approximation, Cvetic and collaborators established a
number of remarkable results.  Perhaps most strikingly, the BPS tension
is the crossover point between two radically different kinds of behavior
of domain wall spacetimes, and coincides with the Coleman-DeLuccia
\cite{cd} bound on the tension of bubbles created by vacuum tunneling.
Coleman and DeLuccia showed that, as a consequence of gravitational
 effects, tunneling into a lower energy vacuum did not always occur.
In AdS space, volume scales like area and so the flat space argument
 that sufficiently large bubbles of lower energy vacuum always grow, 
fails.  Bubbles will only grow if the tension in their walls is below a
 certain value, which Cvetic {\it et. al.} show coincides with the BPS
 tension .   This is a terribly interesting result, whose significance
 has probably not yet been fully understood.

For tension below the BPS bound, domain walls are not static, but
instead correspond to the infinite radius limit of expanding bubbles of
false vacuum.  Similarly, for tensions greater than the BPS bound, there
are no static solutions.  Rather, if we imagine one of the vacua to be
asymptotically flat, then the solution looks like a domain wall with
constant acceleration, which comes in from infinity and then accelerates
away .  It spends most of its history moving with respect to the
observer in the asymptotically flat region with speed almost equal to
the speed of light.  Cvetic {\it et. al.} describe this situation by
saying that both sides of the wall seem to be living on {\it the inside}
of an expanding bubble of false vacuum.  I think instead that 
these solutions are simply providing evidence that an asymptotically
flat observer cannot prepare states that correspond to arbitrarily large
bubbles of the other vacuum.  The would be infinite radius limits of
such bubbles do not admit deformations that would allow us to embed them
in asymptotically flat space.

For the purposes of the present paper, we are interested in pairs of
vacua which both have vanishing cosmological constant.  In this case the
BPS tension goes to zero and the domain walls behave as described above.
We do not expect large bubbles in this case either.

Before leaving this section I should note that Cvetic {\it et. al.} also
investigated models with a dilaton field with a potential which had no
minimum for asymptotic values of the field.  They generically find naked
singularities on the side of the domain wall where the dilaton can run.
This investigation was not carried out for the most general form of
dilaton potential one could imagine in the asymptotic region.  It bears
on the question of whether isolated vacua are in the same theory as
weakly coupled string regions (so that one could hope to gain
information about them by resumming string perturbation theory), and
deserves further investigation.

\section{\bf The trouble with bubbles}

After these preliminary considerations, it is time to set up the main
problem to be solved to resolve the question under study.  We will
consider a Lagrangian with only a single scalar field coupled to
gravity. Multiple field Lagrangians, with generic metric on
field space, and the same scaling properties as the Lagrangian below, 
will behave in exactly the same way. The Lagrangian is
\eqn{lag}{{\cal L} = - \sqrt{-g} [M_P^2 R + (\nabla\phi )^2 + M^4
V(\phi / m)].}

We will assume $M \ll M_P$ to assure the validity of effective field
theory. Lagrangians of this type can be derived for moduli in various
regions of M-theory moduli space.  For example, in Horava-Witten like
compactifications \cite{horwit} we can take $M$ to be related to the eleven
dimensional Planck scale by $M^4 = {(M_P^{11} )^6 \over M_P^2}$.  
$m$ would then be $M_P$ or $M_P^{11}$ depending on whether $\phi$ was a
bulk or boundary modulus\footnote{This parametrization may be
misleading, in that it holds only in the region of moduli space where
the internal volume is large. At the end of this section we will argue
that the barrier between a realistic Horava-Witten vacuum and other
isolated vacua is of order the Planck scale.}.  In weakly coupled heterotic 
string theory with Kahler or racetrack stabilization, we might imagine 
$M$ to be calculated in terms of the string scale and coupling 
as some nonperturbative effect, $M \sim m_s e^{- c/g^2}$ while $m$ was
of order the string scale.  Note that in this case one would, for
phenomenological reasons, want the string and Planck scales to be very
close to each other.  A variety of other possible choices for scales
could be achieved in brane world scenarios with low fundamental scale.

Before beginning our general analysis, let me note that if $m \sim M_P$
, as seems plausible in a variety of contexts, then one can rescale the
variables in the Lagrangian in such a way that the equations become
independent of all parameters.  In such systems there is really only a
single scale, $M^2 /M_P$.  Thus, if our contention that ``otherwordly''
bubbles have a maximum size and lifetime is valid, it implies that the 
order of magnitude of the maximum size is this universal scale.  Thus for bulk
moduli in Horava-Witten compactifications, the maximal size of bubbles
is of order $M_P^2 /(M_P^{11} )^3$ or about $10^9$ times the Planck
scale.  This is also the order of magnitude of their lifetime, and it is
hard to imagine ever distinguishing them from all of the other highly
unstable states in such a theory.

It is worth inserting a short comment on energy scales in string theory
at this point.  Indeed, the argument above appears to be much stronger
for cases where the barriers between vacua are of order the Planck
scale.  
In the Horava-Witten case one is faced with the question of what the
{\it qualitative} difference is between bubbles which are $10^9$ times
the Planck scale and those which are $10^{10}$ kilometers in size.  I
think the correct answer to this is that we really have no understanding
of why any scale in string theory should be much smaller than the Planck
scale. Our penchant for thinking otherwise is based on experiment.
In cases with a moduli space, one can always get low energy scales by
going to boundaries of moduli space.   But it is very difficult and
may be impossible to find an isolated vacuum in such a regime.

The other generic idea for producing low energy scales involves
marginally relevant operators in effective field theory.  A modestly
small deviation from a fixed point along a marginally relevant direction
can generate exponentially small energy scales. The two classes of
extant examples in M-theory involve four dimensional models with either
eight or four supercharges.  In the former case there seems to be no
reason for the gauge theory energy scale to have anything to do with the
size of the barrier between M-theory vacua.  The gauge theory does not
produce a potential on the moduli space.   When there are only four
supercharges we often generate a small potential on the moduli space but the
cosmological constant is generically nonvanishing and we are not in the
class of asymptotically flat vacua we have been investigating in this paper.
While there are supersymmetric gauge theories which generate a small
superpotential and have isolated vacua with R-symmetry at the
minimum\footnote{I thank A. Nelson for pointing out a large class of
examples.}, it is not clear that any of them are realized in string theory. 

The question of scales takes on a new light in view of the conjectures
presented in \cite{lambdatwo}.  There are two classes of low energy
scales that are indicated by experiment.  The first is the collection of
scales associated with standard model physics.  It is plausible that all
of these are related to the SUSY breaking scale, and if the conjecture
of \cite{lambdatwo} is correct, they would not be present in a
limiting, asymptotically flat supersymmetric theory of the type we are
studying here.   The other class are the three pieces of evidence
(coupling unification, neutrino mass, and amplitude of the cosmic
microwave background fluctuations) for a scale a few orders of magnitude
below the Planck scale, which is unrelated to SUSY breaking.   In
Horava-Witten compactifications, this scale is related to a small
potential on moduli space generated by branes in a bulk with eight
supercharges.  However, the small value of the potential is itself
attributed to the fact that the minimum is in an extreme region of
moduli space where the volume of the internal space is large in Planck
units.  It is not easy to understand (particularly in the Horava-Witten
context where the codimension of the standard model brane is $1$) how
the system could have a minimum in this regime.  Thus, it is not
implausible that other isolated minima are in fact separated from the
one describing the real world by barriers of order the (relevant) 
Planck scale.   

\subsection{Properties of bubble solutions}

The investigation of spherically symmetric bubble solutions proceeds by
introducing an ansatz for the metric and scalar field in terms of
angular coordinates $\Omega$ and a two dimensional, radius/time space
$z^a = (t,r)$.  The metric is
\eqn{metric}{ds^2 = g_{ab} (z) dz^a dz^b + \rho^2 (z) d\Omega^2 .}
and $\phi$ is taken to be a function only of $z$.  The equations of
motion may be derived from the two dimensional Lagrangian
\eqn{2dlag}{- \sqrt{-g}[M_P^2 \rho^2 R + M_P^2 (\nabla\rho )^2  +
\rho^2 (\nabla\phi )^2 + M^4 \rho^2 V(\phi /m)].}
The stress tensor of the matter field is
\eqn{stress}{T_{ab} = \partial_a \phi \partial_b \phi - g_{ab}(-{1\over
2} (\nabla\phi )^2 - M^4 V(\phi /m)).}
  
We are looking for solutions that have the following properties at late
times:  

\begin{itemize}

\item There is an inner region where $\phi$ is approximately constant
and equal to one of the isolated minima of $V$.

\item The metric is asymptotically flat and $\phi$ asymptotes to some
other minimum of $V$.

\item The variation of $\phi $ between the two minima takes place
principally in a region of geodesic size $l$.  We will call this the
bubble wall.

\item There is a moment of maximal expansion of the bubble wall, when
all fields are instantaneously static.

\end{itemize}

The motivation for the third property, as well as determination of the
length $l$ comes from the standard energetics of domain walls.  The
distance between the two minima in field space is of order $m$.  Thus,
the energy of the wall at radius $R$ is approximately
\eqn{energy}{E \sim 4\pi R^2 [(m^2 / l) + l M^4 ].} 
This is minimized at $l \sim {m \over M^2}$, giving a tension $T \sim m
M^2$. 

Reserving a detailed investigation of these equations of motion for
future work, we borrow an old result of W. Israel \cite{israel}.  Israel
studied bubbles whose walls were made of dust {\it i.e.} matter with
vanishing pressure.  Though this is not necessarily a good approximation to 
a scalar field stress tensor in the present circumstances, I believe it
captures the essential physics.   Israel worked in a coordinate system
whose radial coordinate is just the field $\rho$. He took the bubble to
be infinitely thin and parametrized its position by a function $R(t)$
where $t$ is proper time in the comoving frame of the bubble wall.
The metric outside the wall is, by Birkhoff's theorem, the Schwarzchild
metric with mass ${\cal M}$.  In our system ${\cal M} = 4\pi R_{max}^2 
T$, where $R_{max}$ is the maximal value of the radius and $T$ is the
tension estimated above.

The equation of motion of the bubble wall is
\eqn{eom}{{\cal M} = \mu (\sqrt{1 + \dot{R}^2} - 1/M_P R ).}
where $\mu$ is an integration constant. This determines 
$\mu = {\cal M}/(1 - 1/R_{max})$.

The problem comes in the consistency condition that $R$ remain outside
the Schwarzchild radius of the external metric. This reads
\eqn{consist}{R > 2 {\cal M}/M_P^2 = 8 \pi T R_{max}^2 / M_P^2}
{\it A fortiori} this must be satisfied at $R_{max}$ so we obtain an
order of magnitude bound
\eqn{bound}{R_{max} < {M_P^2 \over M^2 m},} 
which is the result announced in the introduction.  The introduction of
a more realistic model of the scalar stress tensor should not change
this result dramatically.

We conclude then that an asymptotic observer who tries to create too
large a bubble of otherwordly vacuum will instead create a black hole.
Note that for domain wall tensions of order $1$ TeV the maximum size
bubble is of order $10^{10} km$, so discussions of multiple vacua in low
energy quantum field theory are not substantially effected by our bound.
However, the size shrinks rapidly with rising energy scale for the
domain wall tension, and once we are in the range of unification scales
the gravitational corrections are significant.

\section{Discussion}

There are, I believe, two possible interpretations of the results of
this paper.  The most straightforward is the one advertised in the
introduction .  Isolated asymptotically flat M-theory vacua, and more
generally, disconnected components of the moduli space of vacua with
four dimensional $N = 1$ SUSY are truly in different theories. One cannot
access one from the other.

One might, as in field theory, hope to evade this conclusion by
examining the high energy behavior of the theory, but here again black
holes and the UV/IR connection get in the way.  High energy scattering
in M-theory is almost certainly dominated by black hole production and
decay \cite{vbf}.  Black holes decay by Hawking radiation and the
Hawking temperature goes to zero with the mass.  Thus, by examining the
final states in high energy scattering processes, one probes the low
energy spectrum of the theory.  It is simply not true that high energy
physics becomes independent of the choice of isolated vacuum state.

Note by the way that the same conclusion can not be drawn for the case
of two different points on a continuous moduli space.  The low energy
excitations are the same in this case: massless moduli and gravitons and
gauge fields.  Thus, the asymptotics of high energy black hole
production will be the same at the two points.

Another, more bizarre, interpretation of our results is motivated by the
idea of black hole complementarity \cite{sussetal}.   We have seen that
otherworldly bubbles of too large a size end up inside a black hole.
As part of the explanation of the black hole information paradox,
Susskind and collaborators proposed that the Hilbert space describing
physics as seen by observers falling into a black hole is actually a
part (a tensor factor) of the Hilbert space of external observers .
The great disparity between the description of physics by these two
classes of observers is ascribed to noncommutativity between their
preferred classes of observables.

Could it be that a similar statement applies to the Hilbert spaces of
two isolated vacua: that superpositions of the states of ``our'' world
(if it were asymptotically flat) correspond to states in a completely
different vacuum?  It is not clear what mathematical content such a
statement could have.  Any two infinite dimensional separable Hilbert
spaces are unitarily isomorphic to each other.  So the Hilbert space of
asymptotically flat M theory is the same as that of any quantum
mechanics problem, even one with a single bosonic degree of freedom.  Quantum
theories are distinguished by the differences between their Hamiltonian
operators and other preferred observables.  Once we have fixed those
they are surely unitarily inequivalent for different vacua.

Although I cannot at the moment see a way in which such a
complementarity between different isolated vacua could make sense, 
I can also remember the time when black hole complementarity made no
sense to me.  So perhaps it is worthwhile to keep this crazy idea in the
back of our minds.

The most likely conclusion then is that the quantum moduli space of 
AFM-theory vacua has disconnected components. Different components can
not be viewed a part of the same theory, at least not in any of the ways
familiar to us from classical or quantum field theory. The components are
partially classified by the number of asymptotically flat dimensions and
the number of supercharges.  For eight or more supercharges, it is
plausible (but not proven) that the moduli spaces have only a few
connected components, and that we can even pass from lower dimensional
compactifications to higher dimensional ones by going to the boundaries
of moduli space.  For four supercharges it is likely that there are a
large number of different theories, one for each connected component of
the moduli space.  We may view these as being labelled by expectation
values of scalar fields, but this description has only a limited, low
energy, utility.  We cannot explore regimes of the scalar field potential
near the Planck energy since any attempt to do so leads to the creation
of black holes.  

There are two questions closely related to the issues explored in this
paper, which deserve further study.  The first is the relation between
isolated supersymmetric vacua and extreme regions of moduli space.  Much
of our information about M-theory is gleaned from an examination of weak
coupling string theory or low energy SUGRA, and certain approximate
vacuum states in this regime look tantalizingly similar to the universe
we inhabit.  On the other hand, it is
very likely that the real world does not inhabit such a regime, even in
the limit in which the cosmological constant vanishes and SUSY is
restored.  What then is the relation between perturbative string or
SUGRA vacua, and isolated SUSY minima?  The results of \cite{cvetic}
suggest that domain walls between an isolated minimum and a weak
coupling regime have problematic naked singularities.  It is of great
interest to understand the implications of this result for processes
within a given asymptotically flat theory.

Finally, we come to the question of whether or not there is a component
of the moduli space of AFM-theory with no SUSY whatsoever.
In \cite{lambdatwo} I suggested that there was not, and a primary
motivation for the present paper was the thought that these
considerations might provide a proof of my conjecture.  So far this has
not occurred, but the issue is far from settled.

\acknowledgments
This work was supported in part by the Department of Energy, grant
DOE DE-FG02-96ER40959.  The author thanks E. Silverstein and
A. Rajaraman for discussions.  Special thanks go to M. Cvetic for
generously explaining numerous details about her work on domain walls in
supergravity. I thank E.Witten, B. Kol, L. Motl and C. Roemelsberger for
comments which led to most of the changes in this version of the paper.



\newpage
\begin{thebibliography}{19}        

\bibitem{lambdatwo} T.~Banks, {\it QuantuMechanics and CosMology}, Talk
given at the festschrift for L. Susskind, Stanford University, May 2000;
{\it Cosmological Breaking of Supersymmetry?}, Talk Given at Strings
2000, Ann Arbor, MI, hep-th/0007146.

\bibitem{bdcoping} T.~Banks, M.~Dine {\it Coping with strongly coupled
string theory}, Phys. Rev. D50, (1994), 7454 hep-th/9406132; {\it
Quantum moduli spaces of $N=1$ string theories}, Phys. Rev. D53, (1996),
5790, hep-th/9508071.

\bibitem{gf} A.~Guth, E.~Farhi {\it An obstacle to creating a universe
in the laboratory}, Phys. Lett. B183, (1987), 149.

\bibitem{cvetic} M.~Cvetic, S.~Griffies, S.J.~Rey, Nucl. Phys. B381,
(1992), 301; M.~Cvetic, S.~Griffies, Phys. Lett. B285, (1992), 27;
M.~Cvetic, S.~Griffies, {\it Domain Walls in N=1 Supergravity}, in
Proceedings of the International Symposium on Black Holes, Membranes and
Wormholes, The Woodlands, Texas, January 1992, edited by S.Kalara and
D.V. Nanopoulos, World Scientific, Singapore 1992.
M.~Cvetic, S.~Griffies, H.~Soleng, {\it Nonextreme and
ultrextreme domain walls and their global space-times},
Phys. Rev. Lett., 71, (1993), 670; {\it Local and global gravitational
aspects of domain wall space-times}, Phys. Rev. D48, (1993), 2613;
M.~Cvetic, H.~Soleng, {\it Naked Singularities in dilatonic domain wall
space-times}, Phys. Rev. D51, (1995), 5768.

\bibitem{cd} S.~Coleman, F.~De Luccia, Phys. Rev. D21, (1980), 3305.


\bibitem{banksah}O.~Aharony, T.~Banks, {\it Note on the quantum
mechanics of m-theory}, JHEP, 9903, 016, (1999), hep-th/9812237.

\bibitem{bfbh} T.~Banks, W.~Fischler {\it A model for
high energy scattering in quantum gravity}, hep-th/9906038.

\bibitem{vbf} H.~Matschull, {\it Black hole creation in 2+1-dimensions}
Class. Quant. Grav. 16, 1069 (1999), gr-qc/9809087;
D.~Amati, M.~Ciafaloni and G.~Veneziano,
{\it Superstring collisions at planckian energies}, Phys. Lett.  B197, 81 (1987),
{\it Can Space-Time Be Probed Below The String Size?}
Phys. Lett. B216, 41 (1989),
{\it Planckian scattering beyond the semiclassical approximation},
Phys. Lett. B289, 87 (1992),
{\it Effective action and all order gravitational eikonal at Planckian energies}
Nucl. Phys. B403, 707 (1993); T.~Banks, W.~Fischler, {\it A model for
high energy scattering in quantum gravity}, hep-th/9906038.

\bibitem{kap} A.~Kapustin, {\it On the universality class of little
string theories}, hep-th/9912044.

\bibitem{sw} L.~Susskind, E.~Witten, {\it The holographic bound in
Anti-DeSitter space}, hep-th/9805114.

\bibitem{horwit} P.~Horava, E.~Witten, {\it Heterotic and Type I string
dynamics from eleven dimensions}, Nucl. Phys. B460, (1996),506, hep-th/9510209; {\it
Eleven dimensional supergravity on a manifold with boundary},
Nucl. Phys. B475, (1996), 94, hep-th/9603142 ; E.~Witten, {\it Strong
coupling expansion of Calabi-Yau compactification}, Nucl. Phys. B471,
(1996), 135.

\bibitem{israel} W.~Israel, Nuovo Cimento 44B, (1966), 1;
Phys. Rev. 153, (1967) 1388.

\bibitem{sussetal} L.~Susskind, L.~Thorlacius, J.~Uglum, {\it The
stretched horizon and black hole complementarity}, Phys. Rev. D48,
(1993), 3743, hep-th/9306069.


\end{thebibliography}
\end{document}